\documentclass[reprint]{jpp}

\usepackage{graphicx}
\usepackage{subfigure}

\usepackage{epstopdf, epsfig}
\usepackage{hyperref}
\usepackage{amsmath,amsfonts,amssymb}
\usepackage{mathptmx}
\usepackage{dcolumn}	% Align table columns on decimal point
\usepackage{bm}			% bold math
\usepackage{etoolbox}
\usepackage{lipsum}
\usepackage{xcolor}

\usepackage{times}

\shorttitle{Plasma Fluctuations in Relativistic Flows}
\shortauthor{A. G. Tevzadze}
\title{Spectral Universality of Turbulent Fluctuations in Relativistic Flows}

\author{A. G. Tevzadze\aff{1,2}}

\affiliation{
\aff{1}Evgeni Kharadze Georgian National Astrophysical Observatory, Abastumani 0301, Georgia
\aff{2}Ivane Javakhishvili Tbilisi State University, Tbilisi 0179, Georgia} 

\begin{document}

\maketitle

\begin{abstract}
We develop a Lorentz-covariant framework for projecting spacetime spectra into temporal spectra of stationary turbulent fluctuations in relativistic flows. For self-similar spacetime spectra, we derive a universal scaling relation, $\alpha = \beta - D$, where $\alpha$ is the temporal spectral index, $\beta$ the spacetime homogeneity exponent, and $D$ the effective dimensionality of spectral support. We further demonstrate that this universality breaks down when spacetime homogeneity is violated. Temporal spectra in relativistic flows are thus intrinsically nonlocal observables, requiring a covariant projection framework that establishes a general principle for spectral inference in relativistic plasma turbulence and high-energy plasma flows.
\end{abstract}

%\keywords{Anisotropic Plasma, Plasma Invariants}

\section{Introduction}

Turbulence and stochastic fluctuations are fundamental features of plasma flows in both laboratory and astrophysical settings. In many such systems, including magnetized relativistic outflows and collisionless plasma turbulence, fluctuations are embedded in flows with bulk relativistic velocities 
(see \cite{Abdo2010,Marscher2008}). Numerical simulations further indicate the presence of broadband spectral structure in relativistic plasma turbulence (see \cite{Schekochihin2022,ComissoSironi2020,Beresnyak2019,Brandenburg2017,ZrakeMacFadyen2012}). In this context, statistical descriptions of fluctuations play a central role, with spectral analysis providing a primary quantitative framework for characterizing the scale dependent variability of random fields (see \cite{Alexakis2018,Frisch1995}).
Spectral quantities are naturally related to correlation functions through the Wiener--Khinchin theorem (see \cite{Wiener1930,Khinchin1934,MoninYaglom1975}), establishing a direct link between observable fluctuations and the underlying statistical structure of the stochastic field.

Within this spectral framework, the temporal spectra obtained from observations depend on the measurement procedure, a distinction well known in turbulence, where Eulerian and Lagrangian frequency spectra probe different aspects of the same stochastic process (see \cite{TennekesLumley1972,Kaneda1993,Biferale2005,Pope2000}). Measurements at a fixed point mix spatial and temporal variations, whereas sampling along trajectories follows the intrinsic evolution of the flow (see \cite{Tennekes1975,Muller2012,Biferale2019}). In general, these descriptions are not equivalent, and the relation between temporal and spatial spectra depends on the sampling geometry. 
In the regime where intrinsic temporal variations of the stochastic field are negligible compared to the advection timescales, a reduced description becomes possible. In this regime, and in the absence of specific spectral features, a widely used identification between temporal and spatial spectra is provided by Taylor's hypothesis (see \cite{Taylor1938,TennekesLumley1972}). Under this assumption, a temporal signal at a fixed point becomes equivalent to sampling along a straight trajectory with constant velocity, so that the temporal spectrum directly reflects the underlying spatial spectral structure (see \cite{Matthaeus2010,Linkmann2017}).

In relativistic plasma turbulence and high-energy astrophysical flows, temporal spectra are often the primary observable, making the relation between temporal and spatial structure particularly critical. 
In relativistic settings, where measurements are performed along worldlines, the relation between temporal and spatial spectra can instead be formulated in terms of spacetime structure (e.g., \cite{Boyanovsky2004,Mukhanov1992,ClarkDiLeoni2020}), rather than through advection based assumptions of Taylor's hypothesis. Measurements along worldlines are formally analogous to Lagrangian sampling in turbulence, where fields are evaluated along trajectories. In contrast to the Lagrangian case, where trajectories are determined by the flow dynamics, the worldline acts as an external geometric constraint, and the resulting temporal spectrum arises as a projection of the underlying spacetime spectral density. Temporal spectra measured along single worldlines are widely used to characterize variability and to infer underlying physical structure, particularly in relativistic systems 
(see \cite{Basboll2019,Marscher2008,Uttley2005}).

Still, a temporal signal measured along a worldline does not correspond to a direct sampling of a spatial field, but instead probes a projection of an underlying spacetime stochastic process. Consequently, a direct identification between temporal and spatial spectra is not generally valid. Indeed, the common identification of temporal and spatial spectra, based on the non-relativistic Taylor hypothesis, has been questioned already in classical turbulence due to intrinsic temporal dynamics and properties of the stochastic field (see \cite{Mahlmann2019,Hada2017,He2019,ClarkDiLeoni2020}). In relativistic systems, temporal and spatial fluctuations are not uniquely separated but depend on the observer, suggesting an intrinsically relativistic origin for deviations between their spectra. The frequency measured along a detector’s worldline is not an independent time component but part of a spacetime four-vector whose components mix under Lorentz transformations. Consequently, slope inference routinely applied in relativistic contexts lacks a formulation consistent with relativistic invariance. This raises a fundamental question: 
\emph{What is the relation between temporal and spatial spectra of stochastic fluctuations in relativistic flows?}

As we show below, the absence of a covariant local mapping leads to a systematic bias in interpreting temporal spectra in relativistic flows. This issue reflects the relativistic structure of measurement itself, independent of any specific dynamical model or statistical assumptions. 

In this work, we analyze the temporal power spectral density measured along an inertial worldline as a projection of an underlying spacetime spectral density onto the observer’s four-velocity, and formulate this relation for stationary stochastic scalar fields in geometric terms. We examine how this projection constrains temporal spectral behavior under different scaling assumptions and illustrate the construction using a synthetic spacetime spectrum. The domain of validity of the present formulation includes stationary stochastic scalar fields with a well defined spacetime spectral density, measured by inertial observers, and restricted to regimes where homogeneous scaling holds over a finite range of scales.

The formulation of the spectral projection, its scaling properties, synthetic spectrum examples, and limits of validity are presented in Sec.~II. Conclusions are summarized in Sec.~III.

\section{Spectral Projection and Universality}

The temporal spectrum arises as a projection of the underlying spacetime spectral density.
To compute this projection explicitly, consider a real valued scalar random field $F(x)$ in Minkowski spacetime with coordinates $x^\mu = (c t, \mathbf{r})$. Assuming second order stationarity in spacetime (statistical translation invariance), the two-point autocorrelation function depends only on the separation variable $\xi^\mu$:
\begin{equation}
R(\xi) = \left\langle F(x) ~ F(x + \xi) \right\rangle ~.
\label{eq:Rxi}
\end{equation}
The invariant four-dimensional spectral density of this quantity can be defined via the covariant form of the Wiener-Khinchin transform (see \cite{MoninYaglom1975}):
\begin{equation}
S(k) = \int {\rm d}^4 \xi ~ R(\xi) \exp \left( {\rm i} k_\mu \xi^\mu \right) ~,
\label{eq:WienerKhinchin}
\end{equation}
where the wave four-vector is defined as $k_\mu = (\omega/c, -\mathbf{k})$, and metric signature $(+,-,-,-)$ is adopted. We restrict attention to scalar observables (e.g., intensity or emissivity), so that the two-point correlation function is a Lorentz invariant scalar. 
An observer with four-velocity $u^\mu = \gamma (c, \mathbf{v})$ samples the stochastic field along its worldline, thereby converting spacetime fluctuations into a temporal signal:
\begin{equation}
x^\mu(\tau) = x_0^\mu + u^\mu \tau ~,
\label{eq:worldline}
\end{equation}
where $\tau$ is the observer's proper time. Similarly, the proper time autocovariance along the worldline depends on the proper time lag variable $s$:
\begin{equation}
R_u(s) = \left\langle F(x(\tau)) ~ F(x(\tau + s)) \right\rangle ~.
\label{eq:Ru}
\end{equation}
Hence, the power spectral density (PSD) of the stochastic signal in the observer's frame may be defined as:
\begin{equation}
P_u(\Omega) = \int {\rm d} s ~ R_u(s) \exp \left( {\rm i} \Omega s \right) ~,
\label{eq:PuDef}
\end{equation}
where $\Omega$ denotes the angular frequency measured by the observer.
Measuring separation variable along the worldline:
\begin{equation}
\xi^\mu = x^\mu (\tau + s) - x^\mu(\tau) = u^\mu s ~,
\end{equation}
we may derive: 
\begin{equation}
R_u(s) = R(us) ~.
\label{eq:RuS}
\end{equation}
Using the inverse Fourier transform of the spectral density:
\begin{equation}
R(\xi) = \int \frac{{\rm d}^4 k}{(2 \pi)^4} ~ S(k) \exp \left( - {\rm i} k_\mu \xi^\mu \right) ~,
\label{eq:Rinv}
\end{equation}
and substituting Eqs. \eqref{eq:RuS} and \eqref{eq:Rinv} into Eq. \eqref{eq:PuDef} and integrating over the proper time lag:
\begin{equation}
\int {\rm d} s ~ \exp \left[ {\rm i} (\Omega - k_\mu u^\mu) s \right] = 2 \pi \delta(\Omega - k_\mu u^\mu) ~,
\label{eq:delta}
\end{equation}
we obtain the measured temporal PSD as the Lorentz invariant projection of the spacetime spectrum:
\begin{equation}
P_u(\Omega) = \int \frac{{\rm d}^4 k}{(2 \pi)^3} ~ S(k) ~ \delta(\Omega - k_\mu u^\mu) ~.
\label{eq:PuProj}
\end{equation}
Thus, the temporal PSD measured by an observer with $u^\mu$ is not $S(\omega)$ but the Lorentz-invariant projection defined by the Eq. \eqref{eq:PuProj} of the full spacetime spectrum onto the hyperplane $k_\mu u^\mu=\Omega$. Hence, the observed angular frequency is defined by the construct:
\begin{equation}
\Omega = \gamma (\omega - \mathbf{v} \cdot \mathbf{k}) ~,
\end{equation}
rather than the frame dependent frequency component $\omega$, indicating that the observed frequency is not an independent temporal variable but part of a Lorentz mixed spacetime form.

Although physical spacetime is 4-dimensional, the stochastic fields need not to occupy the full 3D spatial momentum space. In many systems, fluctuations may be effectively confined to lower dimensional structures, such as planar ($2$D), filamentary ($1$D), or strongly anisotropic configurations, so that the spectrum is concentrated on a lower-dimensional subset of $\mathbf{k}$-space. We therefore denote by $D \le 3$ the effective spatial dimension of the spectral manifold in momentum space, defined as the dimensionality of the region in $\mathbf{k}$-space over which the spectral density $S(\omega,\mathbf{k})$ carries significant spectral weight. In general, the invariant constraint $\Omega = k_\mu u^\mu$ intersects this spectral manifold and restricts integration to a $D$-dimensional subset embedded in $(D+1)$-momentum space. The projection formula may therefore be written in the general form:
\begin{equation}
P_u(\Omega) = \int \frac{{\rm d}^{D+1}k}{(2\pi)^D} ~S(k)~ \delta(\Omega-k_\mu u^\mu) ~,
\label{eq:PuProjGeneral}
\end{equation}
where $D$ denotes the effective dimensionality of spectral support in momentum space. In the full case $D = 3$, but it may be smaller if the spectral support is reduced.

Temporal spectral inference is therefore not a one-dimensional slice of the spacetime spectrum, but an integral over a hypersurface in momentum space. Consequently, the observed spectrum reflects a nonlocal averaging over momentum space, rather than any local frequency–wavenumber mapping. This geometric structure in particular is what determines how temporal scaling emerges under spacetime homogeneity.

\subsection{Homogeneous Spectra and Universal Scaling}

Suppose the spacetime spectrum is homogeneous, i.e., self-similar under uniform rescaling of the full four-momentum within a scaling range. This condition captures the usual scale-invariant structure encountered in stochastic systems:
\begin{equation}
S(\lambda k_\mu) = \lambda^{-\beta}S(k_\mu) ~.
\label{eq:LorentzHomo}
\end{equation}
Substituting this into the invariant projection \eqref{eq:PuProjGeneral} determines the temporal scaling uniquely. This homogeneity condition treats temporal and spatial components symmetrically and defines the Lorentz-homogeneous class of stochastic spectra.
Under the rescaling $k_\mu=\lambda q_\mu$ we get:
\begin{equation}
{\rm d}^{D+1}k = \lambda^{D+1} {\rm d}^{D+1}q ~,
\end{equation}
while the delta function projects a factor $\lambda^{-1}$ to the observable frequency:
\begin{equation}
\delta( \Omega - \lambda q_\mu u^\mu) = \lambda^{-1} \delta( \Omega/\lambda - q_\mu u^\mu) ~.
\end{equation}
Hence, for a homogeneous spacetime spectrum, the projection Eq. \eqref{eq:PuProjGeneral} yields:
\begin{equation}
P_u(\Omega) = \lambda^{D-\beta} P_u(\Omega/\lambda) ~.
\end{equation}
Setting $\lambda=\Omega/\Omega_0$ gives a scale-free temporal spectrum:
\begin{equation}
P_u(\Omega)\propto \Omega^{-\alpha} ~,
\end{equation}
where the temporal exponent $\alpha$ is fixed by the homogeneity index and the spatial dimensionality:
\begin{equation}
\alpha = \beta-D ~.
\label{eq:universality}
\end{equation}

Equation \eqref{eq:universality} defines a universality relation: within the class of Lorentz-homogeneous spectra, the temporal scaling exponent is uniquely determined by the spacetime homogeneity index $\beta$ and the effective spatial dimensionality $D$, and is independent of the underlying dynamics, micro physics or the observer’s state of motion.

The subtraction of $D$ reflects the transverse phase-space measure of the invariant projection manifold and is purely geometric in origin. Because the derivation relies only on spacetime stationarity and Lorentz symmetry, the exponent is identical for all inertial observers. Although frequency and wavevector mix under Lorentz transformations, the scaling index remains invariant. This invariance may be interpreted as a spectral isomorphism: the projection preserves scaling exponents across all inertial frames within the Lorentz-homogeneous spectral class.

Hence, we refer to Eq.~\eqref{eq:universality} as \emph{Lorentz-covariant spectral universality}: within this class, all inertial observers infer the same temporal scaling from the same spacetime spectrum.

Interestingly, spectral universality described above requires full spacetime homogeneity. 
When this symmetry is absent, the invariant projection of the Eq. \eqref{eq:PuProjGeneral} cannot, in general, be reduced to evaluation along any one-dimensional momentum trajectory.
Equivalently, no Lorentz-covariant local identification
\begin{equation}
\Omega = \Omega(k)
\end{equation}
may reproduce invariant projection for generic spectra with multidimensional support.
The origin of this irreducibility is geometric. The delta constraint in Eq.~\eqref{eq:PuProjGeneral} restricts integration to a $D$-dimensional manifold embedded in $(D+1)$-momentum space. The temporal spectrum is therefore an integral over a transverse phase space measure that cannot be represented by the value of $S$ along any single momentum trajectory. A one-dimensional slice cannot reproduce this higher-dimensional measure. Exact reduction to a local frequency–wavenumber mapping is possible only for stochastic fields that are confined in one spatial dimension ($D=1$).

Thus, in multidimensional relativistic systems, temporal spectral inference is inherently nonlocal in momentum space. Any local identification between temporal frequency and a single spatial scale violates Lorentz covariance.

\subsection{Equal Time Spatial Spectra}

To express derived result in terms of measurable quantities,  we introduce the equal time spatial spectrum:
\begin{equation}
E_D(k) = \int {\rm d} \omega ~ S(\omega,k) ~,
\end{equation}
which characterizes the distribution of fluctuations across spatial scales in a fixed inertial frame. For Lorentz-homogeneous spectra satisfying 
\begin{equation}
S(\lambda \omega, \lambda k_\mu) = \lambda^{-\beta} S(\omega, k_\mu) ~.
\end{equation}
integration over frequency $\omega$ yields the equal time spatial power spectrum:
\begin{equation}
E_D(k)\propto k^{-p}, \qquad p=\beta-1 ~.
\end{equation}
Combining this result with Eq.~\eqref{eq:universality} eliminates the spacetime spectral scaling exponent $\beta$ and yields a covariant relation between the temporal spectral exponent $\alpha$ measured along a worldline and the spatial spectral exponent $p$ defined in a fixed inertial frame:
\begin{equation}
\alpha = p-(D-1) ~.
\label{eq:alpha_p_relation}
\end{equation}
Equation \eqref{eq:alpha_p_relation} expresses Lorentz-covariant spectral universality directly in terms of measurable spectral slopes. The temporal exponent is thus uniquely determined by the equal-time spatial spectrum, with a geometric offset set by the effective spatial dimensionality. This offset implies that a naive identification of temporal and spatial slopes leads to a systematic bias in relativistic settings.

In three spatial dimensions, for stochastic fields filling three-dimensional space:
\begin{equation}
\alpha = p-2 ~.
\end{equation}
This implies that, in relativistic plasma flows, temporal spectra measured {\it in situ} or inferred from radiative variability cannot be directly identified with equal-time spatial spectra, even when the underlying fluctuations are broadband and approximately scale free.

This mismatch is structural rather than dynamical: it arises from invariant projection geometry and does not depend on turbulence phenomenology, isotropy assumptions, or frozen-flow closures. In this sense, Eq.~\eqref{eq:alpha_p_relation} replaces Taylor-type identifications between temporal and spatial spectra with a Lorentz-covariant relation derived directly from spacetime symmetry. For example, a Kolmogorov spectrum with $p = 5/3$ yields $\alpha = -1/3$, illustrating the systematic bias introduced by relativistic projection.

From a physical perspective, temporal variability is classified as red when low frequencies dominate (infrared-dominated spectra) and as blue when high frequencies dominate (ultraviolet-dominated spectra). The covariant relation \eqref{eq:alpha_p_relation} then identifies a geometric threshold at:
\begin{equation}
p = D-1 ~,
\end{equation}
separating these regimes: $p > D - 1$ corresponds to red spectra, $p < D - 1$ corresponds to blue spectra, and the marginal case $p = D - 1$ yields $\alpha = 0$, corresponding to a flat temporal spectrum.

\subsection{Projection of a Synthetic Spectrum}

As an explicit example, we construct a synthetic spacetime spectrum within the class of self-similar spectra homogeneous under uniform rescaling of four-momentum. This construction represents a generic class of scale-invariant spectra with finite support, motivated by turbulent cascades and relativistic fluctuation spectra, and is not restricted to a specific physical model. Let us define:
\begin{equation}
\kappa \equiv \sqrt{\lvert k_\mu k^\mu \rvert},
\label{eq:kappa_def}
\end{equation}
so that $\kappa$ defines the invariant momentum scale. A scale-free Lorentz-invariant spectrum then takes the form:
\begin{equation}
S_{\mathrm{syn}}(\kappa) = S_0 ~ \psi(\kappa) ~ \kappa^{-\beta} ~,
\label{eq:Ssyn_general}
\end{equation}
where $\beta$ is the homogeneity index and $\psi(\kappa)$ is a window function restricting the spectrum to the finite scaling interval:
\begin{equation}
\psi(\kappa) =
\begin{cases}
1, & \kappa_{\min} \le \kappa \le \kappa_{\max} ~, \\
0, & \text{otherwise} ~.
\end{cases}
\end{equation}
This yields a finite frequency range for the stochastic signal:
\begin{equation}
c ~ \kappa_{\min} \le \Omega \le c ~ \kappa_{\max} ~.
\end{equation}
The synthetic spectrum may be normalized by imposing unit total spectral weight:
\begin{equation}
\int {\rm d}^D k ~ S_{\mathrm{syn}}(\kappa) = 1 ~,
\label{eq:spec_norm}
\end{equation}
from which the normalization constant $S_0$ is obtained:
\begin{equation}
S_0 = \left(\int_{\kappa_{\min}}^{\kappa_{\max}} {\rm d} \kappa ~ \kappa^{D-\beta-1}\right)^{-1} ~.
\label{eq:S0}
\end{equation}
The temporal spectrum measured by an observer follows from the invariant projection Eq. \eqref{eq:PuProjGeneral}, which in the observer rest frame reduces to:
\begin{equation}
P_u(\Omega) = \int \frac{{\rm d}^D k}{(2\pi)^D} S_{\mathrm{syn}}(\kappa) ~,
\label{eq:Pu_syn}
\end{equation}
where $\kappa = \left|\Omega^2/c^2 - k^2\right|^{1/2}$.
Rescaling the integration variable as
\begin{equation}
k = \frac{\Omega}{c} ~ q ~,
\end{equation}
which measures momenta in units set by the observed frequency $\Omega$ we obtain:
\begin{equation}
P_u(\Omega)= P_0 \left(\frac{\Omega}{c}\right)^{D-\beta} ~,
\label{eq:Pu_rescaled}
\end{equation}
where $P_0$ is effectively constant within the scaling range:
\begin{equation}
P_0 = \frac{S_0}{(2\pi)^D}
\int_{\mathcal L(\Omega)} {\rm d} q ~ q^{D-1}|1-q^2|^{-\beta/2} ~,
\end{equation}
and the integration domain is defined by:
\begin{equation}
\mathcal L(\Omega)=
\left\{ q : \kappa_{\min} \le \frac{\Omega}{c}\sqrt{\lvert 1-q^2 \rvert} \le \kappa_{\max} \right\}.
\end{equation}
This expression makes explicit that the temporal signal is obtained by integration over a $D$-dimensional momentum manifold. The cutoffs $\kappa_{\min}$ and $\kappa_{\max}$ regularize the spectrum outside the scaling range and enter only through the overall normalization.

\begin{figure}
\centering
\includegraphics[width= 0.6 \columnwidth]{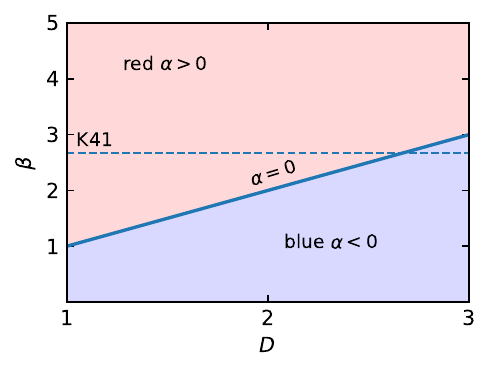}
\caption{Regime map from the spectral universality relation, showing the spacetime spectral index $\beta$ versus the effective dimension $D$ of the stochastic signal. The boundary $\beta = D$ separates blue and red temporal spectra from relativistic projection. The horizontal dashed line marks the Kolmogorov K41 scaling ($\beta_K=8/3$), which projects into the blue regime for $D=3$ and crosses the white spectrum boundary at $D_K \approx 2.67$.}
\label{fig:regime_map}
\end{figure}

The classification of temporal spectra implied by Eq.~\eqref{eq:Pu_rescaled} is summarized in Fig.~\ref{fig:regime_map}, showing the regime structure in the $(\beta, D)$ plane. The geometric threshold $\beta = D$ separates blue ($\alpha < 0$) and red ($\alpha > 0$) temporal spectra, while the boundary itself corresponds to a flat (white) spectrum with $\alpha = 0$. The horizontal dashed line marks the Kolmogorov scaling ($\beta_K = 8/3$). For $D = 3$, this corresponds to a blue temporal spectrum ($\alpha = -1/3$) under invariant projection. The transition to a flat spectrum occurs at the dimension $D_K = 8/3$, where the Kolmogorov line intersects the boundary; beyond this point, the projected spectrum becomes red.

For reference, the temporal spectral indices corresponding to standard turbulence models relevant to plasma flows are summarized in Table~\ref{tab}. For Kolmogorov and Iroshnikov--Kraichnan spectra, the temporal index depends on the effective dimension and may change sign as the dimensionality is reduced, defining a critical dimension $D_*$ at which the temporal spectrum becomes flat and separates blue and red regimes. Thus, the same underlying turbulence spectrum may correspond to different temporal behavior depending on the effective dimensionality. Such reduction may arise from intermittency or anisotropic structure in turbulent plasma flows, where fluctuations are concentrated in sheet-like or filamentary structures in momentum space. In this sense, temporal spectral slopes measured in relativistic plasma systems reflect not only the underlying spatial scaling but also the effective dimensionality of the turbulent cascade. The critical dimension $D_*$ may take non-integer values, consistent with intermittency-reduced filling of phase space (see \cite{Biferale2019,SheLeveque1994,MeneveauSreenivasan1991}).

\begin{table}
\centering
\caption{Examples of spectral indices for common turbulence models and their temporal projections. Here $p$ is the spatial spectral index, while $\alpha_3 = p-2$ and $\alpha_2 = p-1$ are the temporal spectral indices corresponding to three- and two-dimensional spectral support, respectively. The quantity $D_*$ denotes the critical dimension at which the temporal spectrum becomes flat ($\alpha=0$).}
\begin{tabular}{lcccc}
\hline
~ & Spatial index & Temporal index & Temporal index & Critical dimension~ \\
Model & $p$ &   $\alpha_3$ &  $\alpha_2$ & $D_*$ \\
\hline
Kolmogorov (K41) & $5/3$ & $-1/3$ & $2/3$ & $2.67$ \\
Iroshnikov--Kraichnan (IK) & $3/2$ & $-1/2$ & $1/2$ & $2.5$ \\
Spatial White Noise & $1$ & $-1$ & $0$ & $2$ \\
\hline
\end{tabular}
\label{tab}
\end{table}

\subsection{Limits of Spectral Universality}

The universality relation derived above relies on full spacetime homogeneity, which treats temporal and spatial components symmetrically under uniform rescaling. When this symmetry is broken, the invariant projection no longer preserves scaling universally. In the present subsection, we identify the regimes in which this universality breaks down. 
Hence, we consider representative classical spectra to illustrate these regimes.

(i) \textit{Lifshitz-type scaling.}
Classical Lifshitz scaling describes systems in which temporal and spatial directions scale with different weights, characterized by a dynamical exponent $z$ (see \cite{Lifshitz1941,Horava2009}). Such anisotropic scaling arises in turbulent cascades, magnetized plasma turbulence and relativistic perturbation theory. In this class of systems, the spacetime spectrum satisfies:
\begin{equation}
S(\lambda^{z}\omega,\lambda\mathbf{k}) =
\lambda^{-\beta}S(\omega,\mathbf{k}) ~, \qquad z \neq 1 ~.
\end{equation}
Temporal and spatial frequencies then belong to inequivalent scaling sectors. Because the invariant contraction $k_\mu u^\mu$ mixes $\omega$ and $\mathbf{k}$, the projection intertwines these scalings, so that the temporal exponent depends on $z$ and on the observer velocity, and no universal relation between $\alpha$ and $\beta$ is preserved. In this case, explicit calculations accounting for relativistic projection are required.

(ii) \textit{Dispersion dominated spectra.}
In many physical systems, fluctuations are wave-supported rather than broadband, with spectral power concentrated near a dispersion surface $\omega=\omega(\mathbf{k})$, as in wave-supported plasma fluctuations, plasma modes, or relativistic perturbations (see \cite{Schekochihin2022,Beresnyak2019,NewellRumpf2011}). In such cases, the spacetime spectrum takes the form:
\begin{equation}
S(\omega,\mathbf{k}) \sim E_D(\mathbf{k}) ~ \delta \big(\omega-\omega(\mathbf{k})\big) ~,
\end{equation}
and is effectively supported on a lower-dimensional manifold in momentum space.
The invariant projection then reduces to an integral over the intersection of this dispersion surface with the hyperplane $k_\mu u^\mu=\Omega$. The resulting temporal scaling is controlled jointly by the spatial spectrum $E_D(\mathbf{k})$ and the dispersion relation $\omega(\mathbf{k})$. Unlike the Lorentz-homogeneous case, the projection no longer produces a universal dimensional offset; only for linear dispersion does the cancellation leading to $\alpha=\beta-D$ reappear.

Outside the Lorentz-homogeneous class, temporal slopes cannot, in general, be inferred from equal-time spatial exponents alone. The universality relation is therefore symmetry-protected rather than generic. A detailed analysis of anisotropic scaling and dispersion dominated regimes will be presented elsewhere.

\section{Discussion and Summary}

An important question in plasma turbulence with relativistic bulk motion is how temporal variability measured along a single worldline relates to the underlying spatial structure of fluctuations. In relativistic plasma flows, this relation is nontrivial, as temporal and spatial components are not independent but are coupled through the spacetime structure of the measurement.

We demonstrate that temporal spectra measured along timelike worldlines do not provide a direct probe of spatial structure, but instead reflect a projection of the underlying spacetime spectrum. For Lorentz-homogeneous stochastic fields, this projection gives rise to a universal relation between temporal and spatial scaling exponents, whereby the temporal slope is offset from the spatial slope by a geometric factor set by the effective dimensionality of momentum space. This relation is independent of dynamics, observer motion, and microphysics.

Although derived spectral universality relation formally resembles the result obtained under Taylor's frozen flow hypothesis, the underlying mechanism is fundamentally different. Taylor's hypothesis relies on a local identification between frequency and advected wavenumber, effectively reducing the temporal spectrum to a one-dimensional slice of the spatial spectrum. By contrast, the covariant formulation shows that temporal spectra are intrinsically nonlocal, arising from integration over a hypersurface in momentum space.

This geometric origin has direct implications for spectral inference. This correction can be directly applied when interpreting temporal variability in relativistic turbulence, plasma fluctuations, or astrophysical light curves. In three spatial dimensions, temporal slopes systematically underestimate spatial spectral indices by two units. For example, temporal spectra with $\alpha \sim 1~-~2$, as observed in astrophysical red-noise variability (see \cite{Vaughan2003}), correspond to spatial slopes $p \sim 3~-~4$, significantly steeper than would be inferred under naive identification. Thus, the widespread appearance of red temporal spectra in relativistic systems reflects projection geometry rather than universal stochastic driving.

Because the derivation relies only on spacetime stationarity and symmetry, the projection principle applies broadly, including relativistic turbulence and plasma fluctuations (see \cite{Schekochihin2022}), laboratory beam systems (see \cite{HuangKim2007,Chao1993}), and MHD turbulence in early Universe (see \cite{Tevzadze2012}). In all such systems, temporal measurements alone do not uniquely determine spatial structure unless the projection geometry is properly accounted for.

The universality relation is, however, symmetry protected rather than generic. It breaks down when spacetime homogeneity is violated, such as in anisotropic (Lifshitz-type) scaling or dispersion dominated spectra, as encountered in wave supported systems, where temporal and spatial scaling decouple and explicit modeling of the projection becomes necessary.

In this sense, spectral universality defines a well posed regime of relativistic inference: when symmetry holds, temporal scaling follows a universal geometric law; when it does not, temporal spectra must be interpreted through the full spacetime structure. Any consistent extraction of spatial information from temporal observations in relativistic systems must therefore incorporate this covariant projection framework. This establishes a direct, observer independent correction required when inferring spatial structure from temporal measurements in relativistic systems.

\section*{Acknowledgements}

{ChatGPT (OpenAI) was used for language and grammar editing. No scientific content was generated using AI tools.}

\end{document}